\documentclass[10pt,conference]{ieeetran}

\usepackage{listings}
\renewcommand{\thesection}{\Roman{section}}
\renewcommand{\thesubsection}{\arabic{subsection}}

\usepackage{placeins}
\usepackage[T1]{fontenc}
\usepackage[dvipsnames]{xcolor}
\usepackage{amsthm}
\usepackage{thmtools}
\usepackage{mathtools}
\usepackage{braket}
\usepackage{bm} 
\usepackage{bbm} 
\usepackage{dsfont}
\usepackage{amsmath}
\usepackage{amssymb}
\usepackage{enumitem}
\usepackage[colorlinks]{hyperref}
\hypersetup{
    breaklinks  = true,
    colorlinks  = true,
    citecolor   = PineGreen,
    linkcolor   = MidnightBlue,
    urlcolor    = PineGreen
}
\usepackage[utf8]{inputenc} 
\usepackage{tikz}
\usepackage{xurl}
\usepackage{quantikz}

\usepackage{etoolbox}
\apptocmd{\sloppy}{\hbadness 10000\relax}{}{}

\usepackage{caption}
\usepackage[english]{babel}
\usepackage[ruled,linesnumbered]{algorithm2e}
\usepackage{algpseudocode}

\usepackage{graphicx}

\usepackage{pgfplots}
\pgfplotsset{compat=1.18}

\usepackage[acronym]{glossaries}
\glsdisablehyper
\newacronym{qcg}{QCB}{Quantum Circuit Backend}
\newacronym{qft}{QFT}{quantum fourier transformation}

\newcommand{\ptr}[2]{\texttt{#1$\rightsquigarrow$#2}}
\newcommand{\struct}[1]{\texttt{#1}}

\usepackage{layout}

\usepackage[capitalize]{cleveref}
\crefname{figure}{Fig.}{Figs.}
\Crefname{figure}{Figure}{Figures}
\crefname{algocf}{Algorithm}{Algorithms}
\Crefname{algocf}{Algorithm}{Algorithms}

\usepackage[final]{changes}
\definechangesauthor[name={Editor}, color=blue]{ed}

\def\BibTeX{{\rm B\kern-.05em{\sc i\kern-.025em b}\kern-.08em
    T\kern-.1667em\lower.7ex\hbox{E}\kern-.125emX}}

\begin{document}

\title{Speed-oriented quantum circuit backend}
\author{\IEEEauthorblockN{1\textsuperscript{st} Sören Wilkening}
\IEEEauthorblockA{\textit{Institut f\"ur Theoretische Physik} \\
\textit{Leibniz Universit\"at Hannover}, Germany \\
soeren.wilkening@itp.uni-hannover.de}
}

\maketitle

\begin{abstract}
    We present a new software package for efficient quantum circuit generation, designed to achieve optimal runtime performance. Despite being in an early stage of development, our implementation demonstrates significant advantages over existing tools. Using the \replaced[id=ed]{quantum Fourier transform (QFT)}{Quantum Fourier Transform (QFT)} as a benchmark, we show that our \replaced[id=ed]{backend}{algorithm} can generate circuits for systems with up to 2000 qubits faster than widely used frameworks such as Qiskit and Q\#. This improvement is particularly relevant for applications where classical \replaced[id=ed]{preprocessing}{pre-processing} time, including circuit generation, must be minimized to not diminish any potential quantum advantage - for example, in combinatorial optimization tasks. Additionally, our software provides high-level primitives for bit- and integer-level manipulations, offering a simplified interface for integration with high-level quantum programming languages.
\end{abstract}

\begin{IEEEkeywords}
Quantum computing, quantum circuit generation, quantum circuit compilation
\end{IEEEkeywords}

\section{Introduction}

Over the past decade, significant progress has been made in the development of quantum algorithmic frameworks~\cite{advancements}, including quantum algorithms~\cite{algorithms}, hardware architectures, and full-stack quantum computing pipelines~\cite{full-stack}, advancing the field towards achieving practical quantum advantage~\cite{toshio2024practicalquantumadvantagepartially, practical_advantage}. A key component of this effort is the design of high-level quantum programming languages~\cite{Green_2013, silq, binkowski2025cq, faro2025quteshighlevelquantumprogramming, 10.1145/2699415}, which abstract away low-level implementation details and accelerate the discovery and deployment of new quantum algorithms.

The usefulness of such high-level languages critically depends on the availability of efficient backends capable of generating, storing, and optimizing quantum circuits~\cite{rosenhahn2025optimizationdrivenquantumcircuit, QiskitCommunity2017, Cirq_Developers_2025, seidel2024qrispframeworkcompilablehighlevel}. In the near-term era of noisy intermediate-scale quantum (NISQ) devices, optimization often focuses on minimizing qubit counts and gate depth~\cite{Huang_2023, Corcoles2020Challagnes}. However, in the longer term --- when fault-tolerant devices with abundant qubits become available --- other considerations arise. 
In particular, for many applications\added[id=ed]{,} the runtime cost of circuit generation itself becomes a limiting factor, and a backend that emphasizes high throughput while producing circuits with \replaced[id=ed]{``sufficient''}{``sufficient"} levels of optimization becomes essential.

This requirement is especially evident in quantum algorithms for combinatorial optimization~\cite{wilkening2024quantumalgorithmsolving01, Cade2023quantifyinggrover, ammann2023realisticruntimeanalysisquantum, mikuriya2024quantumspeedupquadraticassignment, wilkening2025QuantumSearchMethodQuadratic}, where classical preprocessing, including circuit generation, must not dominate the total runtime. To address this challenge, we present a software package for quantum circuit generation (QCG)\replaced[id=ed]{, introduced in}{that was introduced in}~\cite{Wilkening2026}, which is designed for near-optimal runtime performance. 
Using the quantum Fourier transform (QFT) as a benchmark, we demonstrate that our backend can generate circuits for systems with up to 2000 qubits significantly faster than existing tools.
\replaced[id=ed]{The QFT is chosen as a benchmark because it is a well-known routine that appears in various applications}{The QFT is considered, as it is a well-known routine that appears in various applications}, such as quantum integer arithmetic~\cite{Ruiz_Perez_2017}, quantum amplitude estimation~\cite{Brassard_2002}, and Shor's algorithm~\cite{Shor_1997}, among others.

Beyond basic gate-level construction, our package supports assembly-like instructions for integer and \replaced[id=ed]{Boolean}{boolean} operations, enabling more expressive circuit generation and providing a foundation for high-level quantum programming languages~\cite{Green_2013, silq, binkowski2025cq}. Although the current implementation is an early-stage prototype, it already supports key arithmetic operations such as addition, multiplication~\cite{Ruiz_Perez_2017}, and division~\cite{thapliyal2018quantumcircuitdesignsinteger}, as well as bitwise operations on quantum and hybrid classical--quantum data. 
These capabilities facilitate \replaced[id=ed]{rapid implementation}{a potential rapid implementation} of Grover-type search algorithms \cite{ambainis2004, wilkening2025constraintorientedbiasedquantumsearchforgeneral, wilkening2025constraintorientedbiasedquantumsearch} and other applications that require structured data manipulation.

Our approach contrasts with existing software packages such as Qiskit~\cite{QiskitCommunity2017}, Cirq~\cite{Cirq_Developers_2025}, and others~\cite{Green_2013, Microsoft_Azure_Quantum_Development, amazon-braket-sdk, rosa2025quantumstackketplatform, bergholm2022pennylaneautomaticdifferentiationhybrid, Sivarajah_2020, cheraghi2025ariaquantaquantumsoftwarequantum, Steiger_2018, Killoran2019strawberryfields}, which typically combine circuit generation with simulation or execution features. Our software focuses exclusively on efficient circuit construction, making it easy to integrate with high-level languages, optimizers, or simulators as needed.

In summary, we propose a fast and lightweight backend for quantum circuit generation, designed for future large-scale quantum algorithms. 
By decoupling circuit generation from execution and emphasizing runtime efficiency, our work lays the groundwork for scalable quantum software stacks, supporting the development of more complex, high-level quantum programming paradigms.

\section{Methods}

Due to the desired performance and minimal requirements for data storage, we chose the programming language C~\cite{Kernighan1988TheCP} for our implementation, which has no \replaced[id=ed]{external}{} dependencies. 
For this reason, provided code and data structures \replaced[id=ed]{are}{will be} written in C\deleted[id=ed]{ code}.

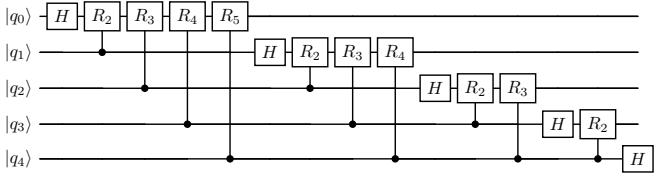
\begin{figure}
    \centering
\resizebox{\linewidth}{!}{%
\begin{quantikz}[sep=4]
\lstick{$\ket{q_0}$} & \gate{H} & \gate{R_2} & \gate{R_3} & \gate{R_4} & \gate{R_5} & \qw      & \qw      & \qw      & \qw      & \qw      & \qw      & \qw      & \qw      & \qw & \qw\\
\lstick{$\ket{q_1}$} & \qw      & \ctrl{-1}    & \qw          & \qw          & \qw          & \gate{H} & \gate{R_2} & \gate{R_3} & \gate{R_4} & \qw      & \qw      & \qw      & \qw      & \qw & \qw\\
\lstick{$\ket{q_2}$} & \qw      & \qw          & \ctrl{-2}    & \qw          & \qw          & \qw      & \ctrl{-1}    & \qw          & \qw          & \gate{H} & \gate{R_2} & \gate{R_3} & \qw      & \qw & \qw\\
\lstick{$\ket{q_3}$} & \qw      & \qw          & \qw          & \ctrl{-3}    & \qw          & \qw      & \qw          & \ctrl{-2}    & \qw          & \qw      & \ctrl{-1}   & \qw &  \gate{H} & \gate{R_2} & \qw \\
\lstick{$\ket{q_4}$} & \qw      & \qw          & \qw          & \qw          & \ctrl{-4}    & \qw      & \qw          & \qw          & \ctrl{-3}    & \qw      & \qw     & \ctrl{-2}     & \qw      & \ctrl{-1}    & \gate{H}
\end{quantikz}
}
\caption{Circuit of a 5-qubit QFT without swaps.}
\label{fig:qft_circuit}
\end{figure}

To store a quantum algorithm, we require certain data structures.
In principle, a quantum gate can be characterized by a few integer and \replaced[id=ed]{floating-point}{float} values, as shown in Data Structure~\ref{dat:gate}.
The minimum information required to describe \replaced[id=ed]{a}{the} gate is its type (\replaced[id=ed]{e.g.,}{like} $X$, $H$, $P$, \ldots), the \replaced[id=ed]{qubit indices it acts on (targets), and the qubit indices it is controlled by (controls)}{qubits/qubit indices it is applied to (targets) and controlled by (controls)}.
\replaced[id=ed]{Parametric gates additionally store the respective parameter value}{Parametric gates also need to store the respective value}.
Some gates \replaced[id=ed]{may}{might} require multiple \replaced[id=ed]{parameters}{parametric values}, but in our description we consider only single\replaced[id=ed]{-parameter}{parameter} gates. 
\replaced[id=ed]{The implementation can easily be extended to support gates with multiple parameters.}{Yet including gates with multiple parameters is straightforward in this implementation.}

\begin{lstlisting}[caption={Structure to store data of a quantum gate.}, label={dat:gate}]
typedef struct gate_t {
    int *controls; // index of control qubits
    int num_controls;
    int *targets; // index of target qubits
    int num_targets;
    int type; // type of gate (X, H, ..)
    double value; // value of parametric gate
} gate_t;
\end{lstlisting}

A straightforward approach to representing a quantum circuit is to store all applied gates sequentially in a one-dimensional list. While this implementation is simple, it becomes inefficient when performing circuit optimization. For instance, consider the case where a newly inserted gate is the inverse of a previously applied gate. Identifying the redundant pair requires searching through the entire list, which incurs a \replaced[id=ed]{time complexity}{time} of $\mathcal{O}(N)$, where $N$ is the current circuit size. 
Moreover, such cancellation is only possible if the two gates occur in consecutive layers, further complicating the optimization process.
To address these limitations, we adopt a layered circuit representation. 
In this model, a layer is defined as a set of mutually independent operations acting on disjoint qubits that can be executed in parallel. 
Organizing gates \replaced[id=ed]{into}{according to} layers enables optimization algorithms to identify and eliminate redundancies locally within each layer, thereby significantly reducing computational overhead.
While a layer-sorted representation can still be implemented using a linear array, this approach would require frequent element rearrangements whenever new gates are inserted, resulting in additional runtime costs. 

\lstset{emph={gate_t}, emphstyle=\color{blue}\bfseries}
\begin{lstlisting}[caption={Structure to store data of a quantum circuit.},label={dat:circuit}]
typedef struct circuit_t {
    gate_t **gates; // [layer][gates]
    int **gate_index; // [layer][qubit]
    int *gates_per_layer;
    int **last_layer_of_qubit; // [qubit][index]
    int *head_last_layer_of_qubit;
    int num_layer;
} circuit_t;
\end{lstlisting}

\begin{figure}
    \centering
    \begin{tikzpicture}
    	\node at (1,5) {gates in layer};
	\node[rotate=-90] at (-.5,4) {layer};
	\draw[->] (-.25,4.75) -- (-.25,3.5);	
	\draw[->] (-.25,4.75) -- (-.25 + 1.25,4.75);	
        \draw[xstep=2.5, ystep=.5,black] (0, 0) grid (7.5,4.5);
        \node at (1.25, .25 + 0.5 * 8) {$H_0$};
        \node at (1.25, .25 + 0.5 * 7) {$CP_{1,0}\left(\frac{\pi}{2}\right)$};
        \node at (1.25, .25 + 0.5 * 6) {$H_1$};
        \node at (3 * 1.25, .25 + 0.5 * 6) {$CP_{2,0}\left(\frac{\pi}{4}\right)$};
        \node at (1.25, .25 + 0.5 * 5) {$CP_{3,0}\left(\frac{\pi}{8}\right)$};
        \node at (3 * 1.25, .25 + 0.5 * 5) {$CP_{2,1}\left(\frac{\pi}{2}\right)$};
        \node at (1.25, .25 + 0.5 * 4) {$H_2$};
        \node at (3 * 1.25, .25 + 0.5 * 4) {$CP_{4,0}\left(\frac{\pi}{16}\right)$};
        \node at (5 * 1.25, .25 + 0.5 * 4) {$CP_{3,1}\left(\frac{\pi}{4}\right)$};
        \node at (1.25, .25 + 0.5 * 3) {$CP_{4,1}\left(\frac{\pi}{8}\right)$};
        \node at (3 * 1.25, .25 + 0.5 * 3) {$CP_{3,2}\left(\frac{\pi}{2}\right)$};
        \node at (1.25, .25 + 0.5 * 2) {$H_3$};
        \node at (3 * 1.25, .25 + 0.5 * 2) {$CP_{4,2}\left(\frac{\pi}{4}\right)$};
        \node at (1.25, .25 + 0.5 * 1) {$CP_{4,3}\left(\frac{\pi}{2}\right)$};
        \node at (1.25, .25 + 0.5 * 0) {$H_4$};
    \end{tikzpicture}
    \caption{Visualization of \replaced[id=ed]{the gates array}{gates array} within the \texttt{circuit\_t} data structure for a \replaced[id=ed]{5-qubit}{5
    five-qubit} QFT without swaps. The \replaced[id=ed]{gate indices}{gates indices} represent their target/control qubits.}
    \label{fig:sequence}
\end{figure}
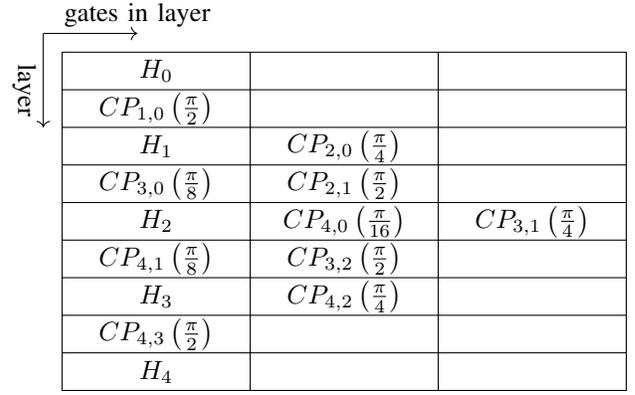

Instead, we employ a two-dimensional data structure (Data Structure~\ref{dat:circuit}), where each row corresponds to a circuit layer and each element within a row stores the gates assigned to that layer. 
This structure enables efficient gate insertion, modification, and cancellation without the need for costly list operations, resulting in $\mathcal{O}(1)$ access and storage time.
An example of the resulting data organization for a 5-qubit \replaced[id=ed]{QFT}{quantum Fourier transform (QFT)} circuit is illustrated in \cref{fig:sequence}. 
Although alternative representations --- such as hash maps keyed by qubit indices or gate identifiers --- could further optimize gate lookup, their exploration lies beyond the scope of this work.

\lstset{emph={circuit_t}, emphstyle=\color{blue}\bfseries}

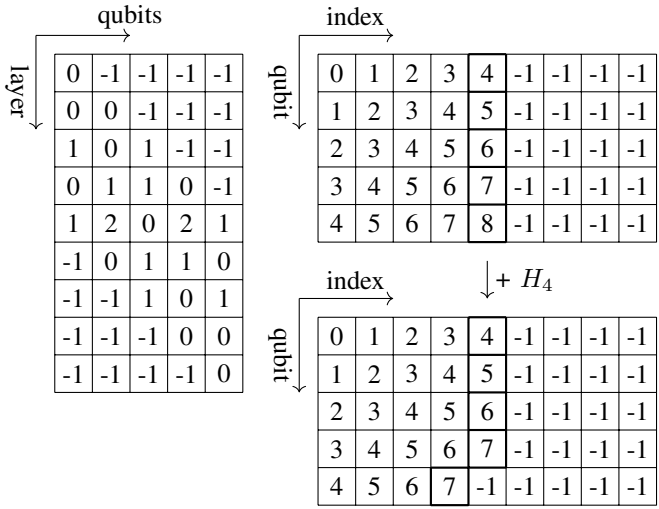
\begin{figure}[t]
    \centering
    \begin{tikzpicture}
    	\node at (1,5) {qubits};
	\node[rotate=-90] at (-.5,4) {layer};
	\draw[->] (-.25,4.75) -- (-.25,3.5);	
	\draw[->] (-.25,4.75) -- (-.25 + 1.25,4.75);	
        \draw[xstep=.5, ystep=.5,black] (0, 0) grid (2.5,4.5);
    	\node at (0.25 + 0.5 * 0, 0.25 + 0.5 * 8) {0};
    	\node at (0.25 + 0.5 * 1, 0.25 + 0.5 * 8) {-1};
    	\node at (0.25 + 0.5 * 2, 0.25 + 0.5 * 8) {-1};
    	\node at (0.25 + 0.5 * 3, 0.25 + 0.5 * 8) {-1};
    	\node at (0.25 + 0.5 * 4, 0.25 + 0.5 * 8) {-1};		
    	\node at (0.25 + 0.5 * 0, 0.25 + 0.5 * 7) {0};
    	\node at (0.25 + 0.5 * 1, 0.25 + 0.5 * 7) {0};
    	\node at (0.25 + 0.5 * 2, 0.25 + 0.5 * 7) {-1};
    	\node at (0.25 + 0.5 * 3, 0.25 + 0.5 * 7) {-1};
    	\node at (0.25 + 0.5 * 4, 0.25 + 0.5 * 7) {-1};		
    	\node at (0.25 + 0.5 * 0, 0.25 + 0.5 * 6) {1};
    	\node at (0.25 + 0.5 * 1, 0.25 + 0.5 * 6) {0};
    	\node at (0.25 + 0.5 * 2, 0.25 + 0.5 * 6) {1};
    	\node at (0.25 + 0.5 * 3, 0.25 + 0.5 * 6) {-1};
    	\node at (0.25 + 0.5 * 4, 0.25 + 0.5 * 6) {-1};		
    	\node at (0.25 + 0.5 * 0, 0.25 + 0.5 * 5) {0};
    	\node at (0.25 + 0.5 * 1, 0.25 + 0.5 * 5) {1};
    	\node at (0.25 + 0.5 * 2, 0.25 + 0.5 * 5) {1};
    	\node at (0.25 + 0.5 * 3, 0.25 + 0.5 * 5) {0};
    	\node at (0.25 + 0.5 * 4, 0.25 + 0.5 * 5) {-1};		
    	\node at (0.25 + 0.5 * 0, 0.25 + 0.5 * 4) {1};
    	\node at (0.25 + 0.5 * 1, 0.25 + 0.5 * 4) {2};
    	\node at (0.25 + 0.5 * 2, 0.25 + 0.5 * 4) {0};
    	\node at (0.25 + 0.5 * 3, 0.25 + 0.5 * 4) {2};
    	\node at (0.25 + 0.5 * 4, 0.25 + 0.5 * 4) {1};		
    	\node at (0.25 + 0.5 * 0, 0.25 + 0.5 * 3) {-1};
    	\node at (0.25 + 0.5 * 1, 0.25 + 0.5 * 3) {0};
    	\node at (0.25 + 0.5 * 2, 0.25 + 0.5 * 3) {1};
    	\node at (0.25 + 0.5 * 3, 0.25 + 0.5 * 3) {1};
    	\node at (0.25 + 0.5 * 4, 0.25 + 0.5 * 3) {0};		
    	\node at (0.25 + 0.5 * 0, 0.25 + 0.5 * 2) {-1};
    	\node at (0.25 + 0.5 * 1, 0.25 + 0.5 * 2) {-1};
    	\node at (0.25 + 0.5 * 2, 0.25 + 0.5 * 2) {1};
    	\node at (0.25 + 0.5 * 3, 0.25 + 0.5 * 2) {0};
    	\node at (0.25 + 0.5 * 4, 0.25 + 0.5 * 2) {1};		
    	\node at (0.25 + 0.5 * 0, 0.25 + 0.5 * 1) {-1};
    	\node at (0.25 + 0.5 * 1, 0.25 + 0.5 * 1) {-1};
    	\node at (0.25 + 0.5 * 2, 0.25 + 0.5 * 1) {-1};
    	\node at (0.25 + 0.5 * 3, 0.25 + 0.5 * 1) {0};
    	\node at (0.25 + 0.5 * 4, 0.25 + 0.5 * 1) {0};		
    	\node at (0.25 + 0.5 * 0, 0.25 + 0.5 * 0) {-1};
    	\node at (0.25 + 0.5 * 1, 0.25 + 0.5 * 0) {-1};
    	\node at (0.25 + 0.5 * 2, 0.25 + 0.5 * 0) {-1};
    	\node at (0.25 + 0.5 * 3, 0.25 + 0.5 * 0) {-1};
    	\node at (0.25 + 0.5 * 4, 0.25 + 0.5 * 0) {0};		

    	\node at (3 + 1,5) {index};
	\node[rotate=-90] at (3.5 + -.5,4) {qubit};
	\draw[->] (3.5 + -.25,4.75) -- (3.5 + -.25,3.5);	
	\draw[->] (3.5 + -.25,4.75) -- (3.5 + -.25 + 1.25,4.75);	
        \draw[xstep=.5, ystep=.5,black] (3.4999 + 0, 1.999) grid (3.5 + 4.5,4.5);
        \draw[xstep=.5, ystep=.5, thick,black] (2 + 3.4999 + 0, 1.999) grid (3.5 + 4.5 - 2,4.5);
    	\node at (3.5 + 0.25 + 0.5 * 0, 0.25 + 0.5 * 8) {0};
    	\node at (3.5 + 0.25 + 0.5 * 1, 0.25 + 0.5 * 8) {1};
    	\node at (3.5 + 0.25 + 0.5 * 2, 0.25 + 0.5 * 8) {2};
    	\node at (3.5 + 0.25 + 0.5 * 3, 0.25 + 0.5 * 8) {3};
    	\node at (3.5 + 0.25 + 0.5 * 4, 0.25 + 0.5 * 8) {4};	
	\node at (3.5 + 0.25 + 0.5 * 5, 0.25 + 0.5 * 8) {-1};		
	\node at (3.5 + 0.25 + 0.5 * 6, 0.25 + 0.5 * 8) {-1};		
	\node at (3.5 + 0.25 + 0.5 * 7, 0.25 + 0.5 * 8) {-1};		
	\node at (3.5 + 0.25 + 0.5 * 8, 0.25 + 0.5 * 8) {-1};				
    	\node at (3.5 + 0.25 + 0.5 * 0, 0.25 + 0.5 * 7) {1};
    	\node at (3.5 + 0.25 + 0.5 * 1, 0.25 + 0.5 * 7) {2};
    	\node at (3.5 + 0.25 + 0.5 * 2, 0.25 + 0.5 * 7) {3};
    	\node at (3.5 + 0.25 + 0.5 * 3, 0.25 + 0.5 * 7) {4};
    	\node at (3.5 + 0.25 + 0.5 * 4, 0.25 + 0.5 * 7) {5};	
	\node at (3.5 + 0.25 + 0.5 * 5, 0.25 + 0.5 * 7) {-1};		
	\node at (3.5 + 0.25 + 0.5 * 6, 0.25 + 0.5 * 7) {-1};		
	\node at (3.5 + 0.25 + 0.5 * 7, 0.25 + 0.5 * 7) {-1};		
	\node at (3.5 + 0.25 + 0.5 * 8, 0.25 + 0.5 * 7) {-1};			
    	\node at (3.5 + 0.25 + 0.5 * 0, 0.25 + 0.5 * 6) {2};
    	\node at (3.5 + 0.25 + 0.5 * 1, 0.25 + 0.5 * 6) {3};
    	\node at (3.5 + 0.25 + 0.5 * 2, 0.25 + 0.5 * 6) {4};
    	\node at (3.5 + 0.25 + 0.5 * 3, 0.25 + 0.5 * 6) {5};
    	\node at (3.5 + 0.25 + 0.5 * 4, 0.25 + 0.5 * 6) {6};	
	\node at (3.5 + 0.25 + 0.5 * 5, 0.25 + 0.5 * 6) {-1};		
	\node at (3.5 + 0.25 + 0.5 * 6, 0.25 + 0.5 * 6) {-1};		
	\node at (3.5 + 0.25 + 0.5 * 7, 0.25 + 0.5 * 6) {-1};		
	\node at (3.5 + 0.25 + 0.5 * 8, 0.25 + 0.5 * 6) {-1};		
    	\node at (3.5 + 0.25 + 0.5 * 0, 0.25 + 0.5 * 5) {3};
    	\node at (3.5 + 0.25 + 0.5 * 1, 0.25 + 0.5 * 5) {4};
    	\node at (3.5 + 0.25 + 0.5 * 2, 0.25 + 0.5 * 5) {5};
    	\node at (3.5 + 0.25 + 0.5 * 3, 0.25 + 0.5 * 5) {6};
    	\node at (3.5 + 0.25 + 0.5 * 4, 0.25 + 0.5 * 5) {7};	
	\node at (3.5 + 0.25 + 0.5 * 5, 0.25 + 0.5 * 5) {-1};		
	\node at (3.5 + 0.25 + 0.5 * 6, 0.25 + 0.5 * 5) {-1};		
	\node at (3.5 + 0.25 + 0.5 * 7, 0.25 + 0.5 * 5) {-1};		
	\node at (3.5 + 0.25 + 0.5 * 8, 0.25 + 0.5 * 5) {-1};	
    	\node at (3.5 + 0.25 + 0.5 * 0, 0.25 + 0.5 * 4) {4};
    	\node at (3.5 + 0.25 + 0.5 * 1, 0.25 + 0.5 * 4) {5};
    	\node at (3.5 + 0.25 + 0.5 * 2, 0.25 + 0.5 * 4) {6};
    	\node at (3.5 + 0.25 + 0.5 * 3, 0.25 + 0.5 * 4) {7};
    	\node at (3.5 + 0.25 + 0.5 * 4, 0.25 + 0.5 * 4) {8};	
	\node at (3.5 + 0.25 + 0.5 * 5, 0.25 + 0.5 * 4) {-1};		
	\node at (3.5 + 0.25 + 0.5 * 6, 0.25 + 0.5 * 4) {-1};		
	\node at (3.5 + 0.25 + 0.5 * 7, 0.25 + 0.5 * 4) {-1};		
	\node at (3.5 + 0.25 + 0.5 * 8, 0.25 + 0.5 * 4) {-1};	

    \draw[->] (5.75, 1.75) -- (5.75,1.25);
    \node at (6.25,1.5) {+ $H_4$};
    
	\node at (3 + 1,5 - 3.5) {index};
	\node[rotate=-90] at (3.5 + -.5,4 - 3.5) {qubit};
	\draw[->] (3.5 + -.25,4.75 - 3.5) -- (3.5 + -.25,3.5 - 3.5);	
	\draw[->] (3.5 + -.25,4.75 - 3.5) -- (3.5 + -.25 + 1.25,4.75 - 3.5);	
        \draw[xstep=.5, ystep=.5,black] (3.4999 + 0, 1.999 - 3.5) grid (3.5 + 4.5,4.5 - 3.5);
        \draw[xstep=.5, ystep=.5, thick,black] (2 + 3.4999 , 1.999 - 3) grid (3.5 + 4.5 - 2,4.5 - 3.5);
        \draw[xstep=.5, ystep=.5, thick,black] (2 + 3.4999 - 0.5, 1.999 - 3.5) grid (3.5 + 4 - 2,4.5 - 5.5);        
    	\node at (3.5 + 0.25 + 0.5 * 0, 0.25 + 0.5 * 8 - 3.5) {0};
    	\node at (3.5 + 0.25 + 0.5 * 1, 0.25 + 0.5 * 8 - 3.5) {1};
    	\node at (3.5 + 0.25 + 0.5 * 2, 0.25 + 0.5 * 8 - 3.5) {2};
    	\node at (3.5 + 0.25 + 0.5 * 3, 0.25 + 0.5 * 8 - 3.5) {3};
    	\node at (3.5 + 0.25 + 0.5 * 4, 0.25 + 0.5 * 8 - 3.5) {4};
		\node at (3.5 + 0.25 + 0.5 * 5, 0.25 + 0.5 * 8 - 3.5) {-1};
		\node at (3.5 + 0.25 + 0.5 * 6, 0.25 + 0.5 * 8 - 3.5) {-1};
		\node at (3.5 + 0.25 + 0.5 * 7, 0.25 + 0.5 * 8 - 3.5) {-1};
		\node at (3.5 + 0.25 + 0.5 * 8, 0.25 + 0.5 * 8 - 3.5) {-1};
    	\node at (3.5 + 0.25 + 0.5 * 0, 0.25 + 0.5 * 7 - 3.5) {1};
    	\node at (3.5 + 0.25 + 0.5 * 1, 0.25 + 0.5 * 7 - 3.5) {2};
    	\node at (3.5 + 0.25 + 0.5 * 2, 0.25 + 0.5 * 7 - 3.5) {3};
    	\node at (3.5 + 0.25 + 0.5 * 3, 0.25 + 0.5 * 7 - 3.5) {4};
    	\node at (3.5 + 0.25 + 0.5 * 4, 0.25 + 0.5 * 7 - 3.5) {5};
		\node at (3.5 + 0.25 + 0.5 * 5, 0.25 + 0.5 * 7 - 3.5) {-1};
		\node at (3.5 + 0.25 + 0.5 * 6, 0.25 + 0.5 * 7 - 3.5) {-1};
		\node at (3.5 + 0.25 + 0.5 * 7, 0.25 + 0.5 * 7 - 3.5) {-1};
		\node at (3.5 + 0.25 + 0.5 * 8, 0.25 + 0.5 * 7 - 3.5) {-1};
    	\node at (3.5 + 0.25 + 0.5 * 0, 0.25 + 0.5 * 6 - 3.5) {2};
    	\node at (3.5 + 0.25 + 0.5 * 1, 0.25 + 0.5 * 6 - 3.5) {3};
    	\node at (3.5 + 0.25 + 0.5 * 2, 0.25 + 0.5 * 6 - 3.5) {4};
    	\node at (3.5 + 0.25 + 0.5 * 3, 0.25 + 0.5 * 6 - 3.5) {5};
    	\node at (3.5 + 0.25 + 0.5 * 4, 0.25 + 0.5 * 6 - 3.5) {6};
		\node at (3.5 + 0.25 + 0.5 * 5, 0.25 + 0.5 * 6 - 3.5) {-1};
		\node at (3.5 + 0.25 + 0.5 * 6, 0.25 + 0.5 * 6 - 3.5) {-1};
		\node at (3.5 + 0.25 + 0.5 * 7, 0.25 + 0.5 * 6 - 3.5) {-1};
		\node at (3.5 + 0.25 + 0.5 * 8, 0.25 + 0.5 * 6 - 3.5) {-1};
    	\node at (3.5 + 0.25 + 0.5 * 0, 0.25 + 0.5 * 5 - 3.5) {3};
    	\node at (3.5 + 0.25 + 0.5 * 1, 0.25 + 0.5 * 5 - 3.5) {4};
    	\node at (3.5 + 0.25 + 0.5 * 2, 0.25 + 0.5 * 5 - 3.5) {5};
    	\node at (3.5 + 0.25 + 0.5 * 3, 0.25 + 0.5 * 5 - 3.5) {6};
    	\node at (3.5 + 0.25 + 0.5 * 4, 0.25 + 0.5 * 5 - 3.5) {7};
		\node at (3.5 + 0.25 + 0.5 * 5, 0.25 + 0.5 * 5 - 3.5) {-1};
		\node at (3.5 + 0.25 + 0.5 * 6, 0.25 + 0.5 * 5 - 3.5) {-1};
		\node at (3.5 + 0.25 + 0.5 * 7, 0.25 + 0.5 * 5 - 3.5) {-1};
		\node at (3.5 + 0.25 + 0.5 * 8, 0.25 + 0.5 * 5 - 3.5) {-1};
    	\node at (3.5 + 0.25 + 0.5 * 0, 0.25 + 0.5 * 4 - 3.5) {4};
    	\node at (3.5 + 0.25 + 0.5 * 1, 0.25 + 0.5 * 4 - 3.5) {5};
    	\node at (3.5 + 0.25 + 0.5 * 2, 0.25 + 0.5 * 4 - 3.5) {6};
    	\node at (3.5 + 0.25 + 0.5 * 3, 0.25 + 0.5 * 4 - 3.5) {7};
    	\node at (3.5 + 0.25 + 0.5 * 4, 0.25 + 0.5 * 4 - 3.5) {-1};
		\node at (3.5 + 0.25 + 0.5 * 5, 0.25 + 0.5 * 4 - 3.5) {-1};
		\node at (3.5 + 0.25 + 0.5 * 6, 0.25 + 0.5 * 4 - 3.5) {-1};
		\node at (3.5 + 0.25 + 0.5 * 7, 0.25 + 0.5 * 4 - 3.5) {-1};
		\node at (3.5 + 0.25 + 0.5 * 8, 0.25 + 0.5 * 4 - 3.5) {-1};
    \end{tikzpicture}
    \caption{\label{fig:lookups}Visualization of \texttt{gate\_index} (left) and \texttt{last\_layer\_of\_qubit} (top and bottom right) within the \texttt{circuit\_t} data structure for a 5-qubit QFT without swaps.
    The table on the bottom right shows how the data structure is adjusted after applying an additional Hadamard gate to qubit~4, eliminating the previously applied gate.
    The thick boxes indicate the current head of the list for the respective qubit. 
    Whenever a gate \replaced[id=ed]{is to be}{has to be} applied to specific qubits, its potential layer position is read from \replaced[id=ed]{this array}{the array}.
    }
\end{figure}

\noindent\textbf{Adding gates to the circuit\added[id=ed]{.}}
Adding a gate to the quantum circuit structure requires \replaced[id=ed]{determining}{knowing} the minimal possible layer where it can be \replaced[id=ed]{placed}{defined}. 
This means that for every qubit that the gate acts on or controls, we need to check \replaced[id=ed]{the last layer where any gate acting on the respective qubit was applied}{which is the last layer where any gate was applied, acting on the respective qubit}.
The increment of the largest of all these values \replaced[id=ed]{gives}{is} the smallest layer where the \replaced[id=ed]{new}{added} gate can be \replaced[id=ed]{placed}{applied}.
Finding the minimal possible layer by checking every applied gate\added[id=ed]{ in a layer-sorted array} \deleted[id=ed]{, considering a layer-sorted array,} could be done with a binary search in time $\mathcal{O}(\log_2(N))$.
\replaced[id=ed]{However, this}{Yet this} can be improved to \deleted[id=ed]{a time of }$\mathcal{O}(1)$\added[id=ed]{} using a lookup table (\replaced[id=ed]{named}{here named} \emph{last\_layer\_of\_qubit}), as described in Data Structure~\ref{dat:circuit} and \cref{alg:mpl}, \replaced[id=ed]{which stores}{storing} the \replaced[id=ed]{relevant}{respective} layer for \replaced[id=ed]{each}{a} qubit.
Rather than \replaced[id=ed]{storing}{only storing} only the last layer, we store a list of all layers that are occupied for a given qubit\replaced[id=ed]{, together with an index tracking the number of entries per qubit}{}.
\begin{algorithm}[t]
\caption{Compute the minimal possible layer below $C$ the gate \struct{g} can be applied to:\\\textsc{MinPossibleLayer}}
\label{alg:mpl}
\KwData{\struct{circuit\_t} *\struct{circ}, \struct{gate\_t} *\struct{g}, \struct{int} $C$}
\KwResult{Minimal possible layer\replaced[id=ed]{ the gate can be applied to}{, gate can be applied}}
$M\gets0$\;
\struct{Q} $\gets $ \ptr{g}{targets} + \ptr{g}{controls} \tcp*{Concat arrays}
\For {$q$ in \struct{Q}}{
    \For{$i =$ \ptr{circ}{head\_last\_layer\_of\_qubit$[q] - 1$}; $i\geq 0$; $i$\texttt{++}}{
        \If{\ptr{circ}{last\_layer\_of\_qubit$[q][i]$} $\leq C$}{
            \textbf{break}\;
        }
    }
    \If{$i < 0$}{
        $i\gets0$ \tcp*{index has to be $\geq 0$}
    }
    $m\gets$ \ptr{circ}{last\_layer\_of\_qubit$[q][i]$}\;
    \If{$m > M$}{
        $M\gets m$\;
    }
}
\Return $M$\;
\end{algorithm}
\begin{algorithm}[t]
\caption{Add \replaced[id=ed]{gate}{Gate} to the quantum circuit:\\\textsc{AddGate}}
\label{alg:add}
\KwData{\struct{circuit\_t} *\struct{circ}, \struct{gate\_t} *\struct{g}}
\KwResult{Boolean}
$P\gets$ INT\_MAX\;
\For{$i = 0$; $i < k$; $i\texttt{++}$}{
    $M= $ \Call{MinPossibleLayer}{\struct{circ}, \struct{g}, $P$}\;
    \If{$M=0$}{
        \textbf{break} \tcp{No gate blocking $g$}
    }
    \struct{G} $\gets$ list of gates in layer $M-1$\;
    \For{\struct{g$_2$} in \struct{G}}{
        \If{$[$\struct{g}, \struct{g$_2$}$]\neq 0$}{ \tcp{\replaced[id=ed]{Gates}{gates} don't commute: stop loop}
            $k\gets0$\;
        }
        \If{\struct{g$_2$} $=$ \struct{g$^\dagger$}}{ \tcp{\replaced[id=ed]{Inverse of previous gate}{is inverse to previous gate}}
            remove \struct{g$_2$} from \ptr{circ}{gates$[M - 1]$}\;
            adjust index lists properly\;
            \Return false \tcp*{\replaced[id=ed]{Gate}{gate} not added}
        }
        \If{\struct{g}, \struct{g$_2$} share same qubits \textbf{and} both types are $P$}{
            \ptr{g$_2$}{value} $\gets$ \ptr{g$_2$}{value} + \ptr{g}{value}\;
            \Return false \tcp*{\replaced[id=ed]{Gate}{gate} not added}
        }
    }
}
append \struct{g} to \ptr{circ}{gates$[M]$}\;
adjust index lists properly\;
\Return true \tcp*{\replaced[id=ed]{Gate}{gate} added}
\end{algorithm}
\deleted[id=ed]{An index will store the number of values associated with every qubit.}
Whenever we add \replaced[id=ed]{the inverse of a}{an inverse} gate \replaced[id=ed]{applied}{to a gate applied} in the last circuit layer, \replaced[id=ed]{the two gates cancel out}{these cancel out}, and we can \replaced[id=ed]{simply decrement}{decrement} the index for the respective qubit without \replaced[id=ed]{recomputing}{the need to recompute} the last occupied layer for \replaced[id=ed]{the relevant}{those} qubits.
For the 5-qubit QFT, this lookup table is shown in \cref{fig:lookups} on the right.
Similarly, we \replaced[id=ed]{apply}{can apply} the same optimization technique when consecutively applying parametric gates\replaced[id=ed]{~---~}{,} such as a controlled phase gate\replaced[id=ed]{~---~}{,} with the same controls and targets.
\replaced[id=ed]{In this case, rather than removing}{Here, we don't remove} the \replaced[id=ed]{existing}{already included} gate, \replaced[id=ed]{we}{but} add the parameter of the \replaced[id=ed]{new}{second} gate to \replaced[id=ed]{that of the existing one}{the parameter of the first gate}~\cite{ramacciotti2023simplequantumalgorithmefficiently}.
To \replaced[id=ed]{locate}{find} the respective gate in the previous layer \added[id=ed]{efficiently}, we introduce \replaced[id=ed]{an additional}{yet another} lookup table, namely \emph{gate\_index} in Data Structure~\ref{dat:circuit}, visualized in \cref{fig:lookups} on the left\deleted[id=ed]{, rather than looping over every gate stored for the respective layer}.
\replaced[id=ed]{This two-dimensional array stores, for each layer and qubit, the index of the corresponding gate in the \emph{gates} array.}{The lookup table is a two-dimensional array, storing, given the respective layer, the index of the gate in the \emph{gates} array for the qubits that the gate acts on.}
Using a sparse array could \replaced[id=ed]{further}{} reduce the memory requirement.

Assuming no inverse gate was applied and no further circuit optimization \replaced[id=ed]{is used}{will be used}, the gate \replaced[id=ed]{is appended}{will be added} to the list of gates for the minimal possible layer.
A simple \replaced[id=ed]{optimization}{circuit optimization}, also integrated in \cref{alg:add}, involves checking \replaced[id=ed]{whether}{if} the new gate commutes with the gates \replaced[id=ed]{in}{within} the last possible layer.
If \replaced[id=ed]{so}{yes}, the gate is ``swapped'' with the gates in the minimal possible layer and the same \replaced[id=ed]{procedure}{routine} as described above \replaced[id=ed]{is}{will be} repeated. 
This is a minimal approach to \replaced[id=ed]{reducing}{minimize} the total \replaced[id=ed]{circuit}{} depth\deleted[id=ed]{ of the circuit}.
\replaced[id=ed]{Although line~7 of \cref{alg:add} loops over up to $N$ gates, this}{Even though in line~7, the complete list of gates in the last possible layer is looped over which would result in a time of $\mathcal{O}(N)$, this} can be optimized to \replaced[id=ed]{consider}{only consider} only the gates \replaced[id=ed]{sharing}{that share} any qubit with the added gate.
\replaced[id=ed]{Since}{As} we assume gates \replaced[id=ed]{contain}{to only contain} only a small\added[id=ed]{,} \replaced[id=ed]{constant}{and constant} number of target and control qubits, this \replaced[id=ed]{reduces}{is reduced} to $\mathcal{O}(1)$.
Furthermore, for \replaced[id=ed]{efficiency}{an efficient implementation}, the swap is \replaced[id=ed]{performed}{only performed} only a constant number of times.
While this risks \replaced[id=ed]{missing some optimizations}{an optimized implementation}, we observe\deleted[id=ed]{,} \replaced[id=ed]{that it appears}{it seems} to be sufficient for most algorithms.

Additional functionality that may be required before adding a gate includes decomposing gates into \replaced[id=ed]{a user-defined gate basis}{gates from a user-defined gate base}.
Furthermore, it would be necessary to \replaced[id=ed]{support restricted}{include the possibility of restricting} connectivity~\cite{AbuGhanem_2025, renger2025superconductingqubitresonatorquantumprocessor}, as not all devices \replaced[id=ed]{provide}{may deliver} all-to-all qubit connections\replaced[id=ed]{; however}{, yet} this is beyond the scope of this work.

\noindent\textbf{Quantum \replaced[id=ed]{instructions}{Instructions}\added[id=ed]{.}}
\replaced[id=ed]{Quantum algorithms such as}{Often, quantum algorithms, like} Grover-type algorithms~\cite{mikuriya2024quantumspeedupquadraticassignment, grover1996fastquantummechanicalalgorithm, ambainis2004, wilkening2024quantumalgorithmsolving01, wilkening2025QuantumSearchMethodQuadratic}\replaced[id=ed]{ are often composed of}{, are of the form that} higher-level operations \replaced[id=ed]{such as}{like} quantum integer addition and comparison\deleted[id=ed]{ are applied}.
Assembly-level quantum instruction sequences significantly restrict optimization opportunities because they erase high-level semantic structure (e.g., arithmetic intent and algebraic relationships), limiting transformations to local, gate-level rewrites.
Additionally, the predefined gate sequences corresponding to operations such as consecutive integer additions introduce strong data dependencies (e.g., via shared carry qubits), which limit commutation, parallelization, and higher-level circuit restructuring.
For this reason, \replaced[id=ed]{it is not necessary}{we are not required} to store a complete quantum circuit for every quantum operation. 
\replaced[id=ed]{Instead, individual instruction sequences can be stored}{We can also store individual sequences} and \replaced[id=ed]{applied}{apply them} sequentially as needed. 
This is shown in Data Structure~\ref{dat:instructions}.
\replaced[id=ed]{A}{Here, a} list of quantum circuits is \replaced[id=ed]{maintained}{stored}, where each quantum circuit comprises only the quantum gates for a specific instruction. 
The instruction list could be, \replaced[id=ed]{e.g.}{f.e}, \{ADD, COMP, cSUB, $\dots$\}.

\begin{lstlisting}[caption={Structure to store an instruction list.}, label={dat:instructions}]
typedef struct instruction_list_t {
    circuit_t *instruction;
    int head; // current instruction
} instruction_list_t;
\end{lstlisting}

\begin{figure*}[!tp]
    \centering
    \includegraphics[width=\linewidth]{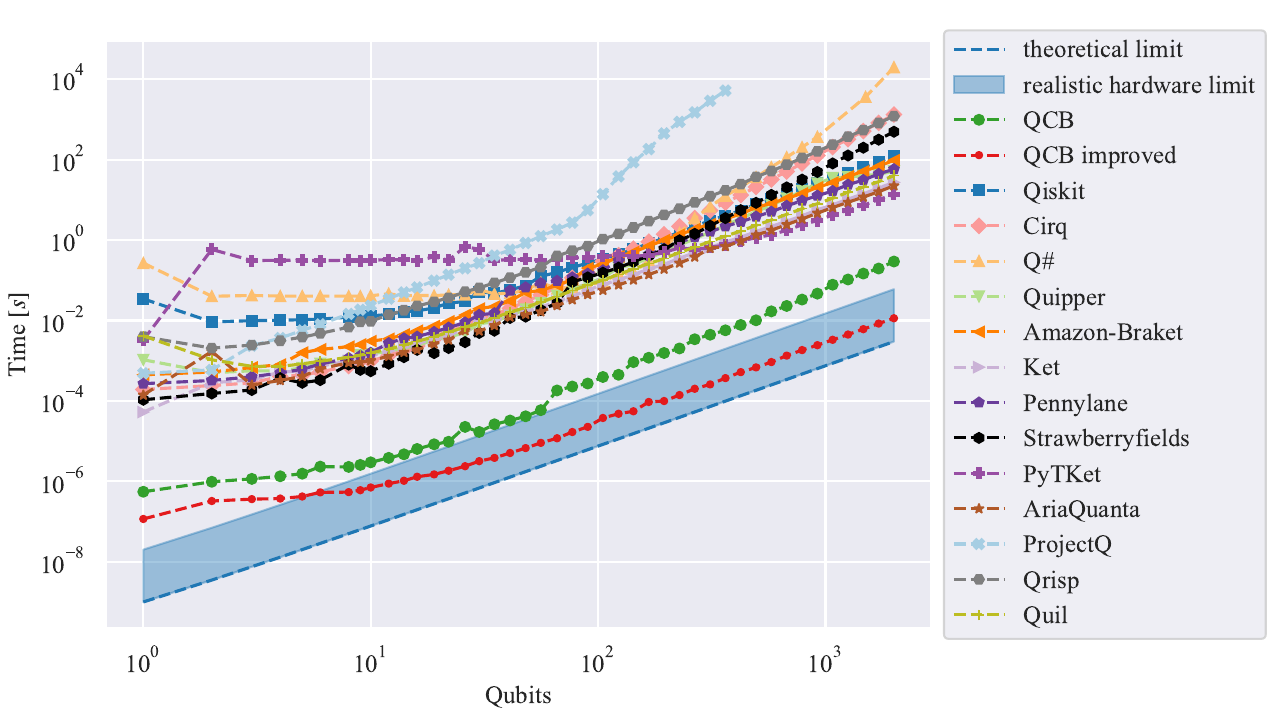}
    \caption{
    	Comparison of running times \replaced[id=ed]{for building}{of state-of-the-art software products to build} the \replaced[id=ed]{QFT circuit across state-of-the-art software packages}{quantum circuit of the QFT}.
	For all instance sizes, our quantum circuit backend generates the \replaced[id=ed]{QFT circuit}{well-known QFT circuit} faster than any other available package.
	Furthermore, the improved \replaced[id=ed]{variant}{quantum circuit backend} approaches the theoretical limit of generating and storing quantum circuits, suggesting that no \replaced[id=ed]{significantly faster single-threaded implementation can be achieved}{faster software can be developed in the future, even if it were single-threaded}.
	}
    \label{fig:runtime}
\end{figure*}

In these cases, we can work with specific, well-known quantum routines, which can be generated more efficiently. 
Rather than applying each gate of the quantum operation individually, \replaced[id=ed]{our software directly adds}{we can implement our software to directly add} each gate to its correct layer. 
A good example of this is the QFT, which also serves as \replaced[id=ed]{a}{the} building block for the quantum Fourier adder~\cite{Ruiz_Perez_2017}.

This type of sequenced quantum instructions also allows for further optimization using tokenization\replaced[id=ed]{~---~}{, }for example, to avoid redundant operations\replaced[id=ed]{}{,} such as \replaced[id=ed]{applying}{checking whether} an inverse QFT \replaced[id=ed]{immediately after}{would be applied after} a QFT.

Additionally, a compiler could determine whether entire operations or code blocks can be executed in parallel without \replaced[id=ed]{verifying}{the need to verify} the layer position for every gate individually.
A mixed strategy with the standard circuit backend can be constructed, such that entire layers rather than \replaced[id=ed]{individual}{} gates are appended to a circuit at the respective smallest possible layer.
\replaced[id=ed]{The instruction list concept is not limited to}{The idea of the instruction list also doesn't not strictly apply only to} well-known, \replaced[id=ed]{preconstructed}{pre-constructed} quantum routines; circuits built using the basic backend can also be stored as a single quantum instruction.

The \replaced[id=ed]{instruction list representation}{representation using an instruction list} also simplifies the incorporation of classical operations and \replaced[id=ed]{classically conditioned}{the classical conditioned} execution of quantum routines, potentially impacting \replaced[id=ed]{algorithm design involving}{the algorithm design of} mid-circuit measurements~\cite{Graham_2023}.

\begin{figure*}[t]
    \centering
    \includegraphics[width=\linewidth]{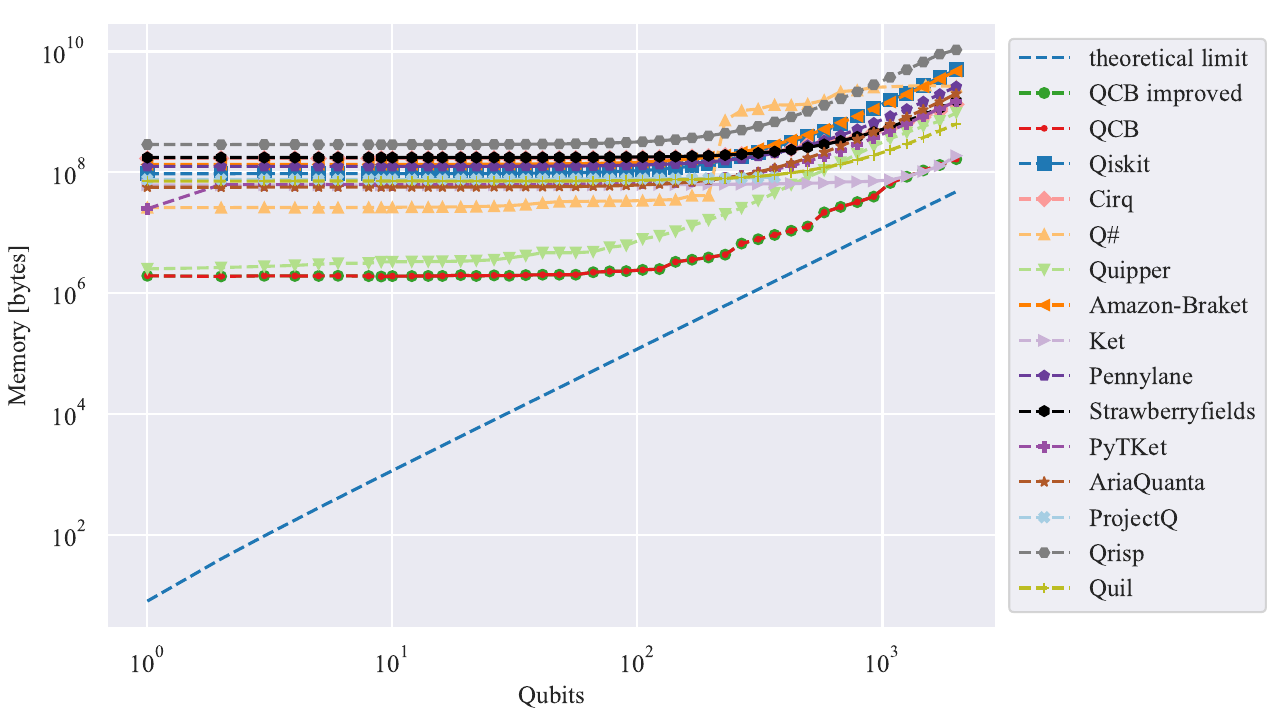}
    \caption{
    	Comparison of \replaced[id=ed]{memory requirements for building}{the memory requirements of state-of-the-art software products to build} the \replaced[id=ed]{QFT circuit across state-of-the-art software packages}{quantum circuit of the QFT}.
	Compared to all \replaced[id=ed]{packages}{solvers} except Ket, our new circuit backend requires \replaced[id=ed]{less memory}{a lower memory requirement}. 
	\replaced[id=ed]{For large instances, Ket achieves a slight advantage; however, we expect this advantage to diminish for larger circuits}{Also, only for large instances, Ket gains a small advantage. 
	We expect Kets' advantage to reduce for larger quantum circuits}, as our algorithm approaches the theoretical \replaced[id=ed]{minimum of the data required for storage}{limit of the minimal data required to store}.
	Both \replaced[id=ed]{variants of our implementation require the same memory, as they use the same underlying data structure.}{of our implementations require the same memory, as they use the}
    }
    \label{fig:memory}
\end{figure*}

\noindent\textbf{Benchmarking\added[id=ed]{.}}
We provide two circuit generation methods\replaced[id=ed]{:}{,} the one where each gate of the QFT is individually added to the circuit structure (\gls{qcg}), and the one treating the QFT as a predefined instruction, storing each gate to its previously known layer (\gls{qcg} improved). 
These two \replaced[id=ed]{methods are}{algorithms will then be} compared \replaced[id=ed]{against}{to} \replaced[id=ed]{12}{11} currently available quantum software \replaced[id=ed]{packages for generating}{generation} quantum circuits, namely Quipper~\cite{Green_2013}, Qiskit~\cite{QiskitCommunity2017}, Cirq~\cite{Cirq_Developers_2025}, Q\#~\cite{Microsoft_Azure_Quantum_Development}, Amazon Braket~\cite{amazon-braket-sdk}, Ket~\cite{rosa2025quantumstackketplatform}, \replaced[id=ed]{PennyLane}{Pennylane}~\cite{bergholm2022pennylaneautomaticdifferentiationhybrid}, TKET~\cite{Sivarajah_2020}, AriaQuanta~\cite{cheraghi2025ariaquantaquantumsoftwarequantum}, ProjectQ~\cite{Steiger_2018}, Strawberry Fields~\cite{Killoran2019strawberryfields} and Qrisp~\cite{seidel2024qrispframeworkcompilablehighlevel}.
Whenever a predefined QFT routine is available\deleted[id=ed]{ to perform the quantum Fourier transformation}, it is used; otherwise, the gate sequence is implemented manually to \replaced[id=ed]{produce}{generate} accurate runtime and memory measurements as efficiently as possible.
The theoretical limit is determined by the minimal amount of data that \replaced[id=ed]{must}{has to} be stored and processed to define the complete QFT.
Here, we assume that every value requires \replaced[id=ed]{16 bits of memory}{16-bit memory}, as no value greater than $2^{16}$ needs to be stored, and can be processed within a single CPU cycle.
As the gates \replaced[id=ed]{must}{have to} be stored within an array, there \replaced[id=ed]{is}{will be} a runtime overhead \replaced[id=ed]{for}{} accessing and storing data within array elements~\cite{drepper2007memory}, resulting in a realistic hardware limit.
\replaced[id=ed]{An}{A} $n$-qubit QFT requires $n$ single-qubit \added[id=ed]{Hadamard }gates and $\frac{n(n-\replaced[id=ed]{1}{3})}{2}$ \replaced[id=ed]{singly}{single} controlled phase gates.
All experiments were conducted on a MacBook Pro (2019) with an \added[id=ed]{Intel }i7 processor, featuring six cores at 2.6\,GHz (4.1\,GHz turbo boost).
The code and \deleted[id=ed]{the performed }experiments are \replaced[id=ed]{available at}{found in}~\cite{Software}.

\section{Results}

As shown in \cref{fig:runtime} and \cref{fig:memory}, both variants of our approach --- the baseline quantum circuit \replaced[id=ed]{backend}{generator} (\gls{qcg}) and the optimized version (\gls{qcg}-improved) --- consistently outperform existing circuit generation frameworks in terms of both runtime and memory consumption.
For small circuits (1--10 qubits), the runtime of our implementations already outperforms the fastest existing tools by 2 to 3 orders of magnitude. 
This improvement holds even for an increasing number of qubits, meaning that the early improvements are not just due to the constant overheads of the existing tools. Specifically, for QFT instances up to 2000 qubits, we observe peak runtime \replaced[id=ed]{improvements}{improvement factors} of up to $47\times$ (\gls{qcg}) and $1209\times$ (\gls{qcg}-improved) compared to PyTKet, and up to $67{,}800\times$ and $1{,}748{,}000\times$, respectively, compared to Q\#. 
Because most existing backends have similar asymptotic complexity, these improvements are expected to persist --- or even grow --- for larger problem sizes.
Importantly, the \gls{qcg}-improved implementation approaches the theoretical lower bound for circuit generation time when constructing \replaced[id=ed]{predefined}{pre-defined} circuits such as the QFT. 
Even the general-purpose \gls{qcg} implementation introduces only a small overhead relative to this optimum.

Similar trends are observed for memory usage. Our approaches require only 100--300\,MB of RAM to generate QFT circuits with 2000 qubits, approaching the theoretical minimum storage required to represent the circuit structure. Among all \replaced[id=ed]{benchmarked tools}{benchmarks}, only Ket slightly outperformed our implementation in terms of memory efficiency for the largest circuits, while all other tools required several gigabytes of memory.

\section{Discussion}
The results demonstrate that our quantum circuit generation backend achieves unprecedented runtime and memory efficiency across a wide range of circuit sizes. Because the \gls{qcg} framework is general and does not rely on assumptions about the specific circuit structure, we expect similar performance gains for arbitrary quantum circuits, not just \replaced[id=ed]{structured}{for structured} examples like the QFT.
These findings have several important implications. First, they \replaced[id=ed]{show}{demonstrate} that classical circuit generation --- often regarded as a negligible preprocessing step --- can\added[id=ed]{,} in fact\added[id=ed]{,} become a significant bottleneck as quantum hardware scales. By reducing this cost to near-optimal levels, our approach enables more accurate estimates of the total wall-clock time of end-to-end quantum workflows. This is particularly relevant for applications such as combinatorial optimization, where classical preprocessing must remain subdominant to preserve any potential quantum speedup.
Second, the substantial memory savings demonstrated here suggest that our backend can scale to very large problem sizes without prohibitive resource requirements. This is a crucial property for integration into cloud-based or distributed quantum software environments, where memory efficiency directly impacts cost and scalability.
Several directions for future work remain. Integrating advanced circuit optimization passes and architecture-aware compilation will allow our backend to target specific hardware constraints while maintaining high generation speed. Incorporating automatic quantum error-correction translation from logical to physical circuits is another key step, both to improve execution fidelity and to provide more realistic performance estimates for large-scale quantum algorithms. Finally, further reducing the constant-time overhead and incorporating multi-threading would make the system even more broadly applicable.
Overall, our results indicate that fast, memory-efficient circuit generation is achievable without sacrificing generality. We believe this approach can serve as a foundational component for future quantum software stacks and high-level programming environments targeting large-scale quantum computation.

\section{Acknowledgments}
This project is \replaced[id=ed]{funded}{founded} by the project ResourceQ.
\replaced[id=ed]{We}{I} thank Lennart Binkowski and Tobias J. Osborne for \replaced[id=ed]{insightful}{very insightful} discussions.

\section{Data and Code availability}
The code and the performed experiments are \replaced[id=ed]{available at}{found in}~\cite{Software}.

\bibliographystyle{ieeetr}
\bibliography{sample}

\end{document}